\documentclass[11pt,a4paper]{article}
\pdfoutput=1
\usepackage[utf8]{inputenc}
\usepackage{amsmath,amssymb,graphicx,epsfig,color,float,cancel,cite}
\usepackage{caption}
\usepackage[compat=1.1.0]{tikz-feynman}
\usepackage{jheppub}
\usepackage{etoolbox}
\newcommand{\h}{\hspace{0.1cm}}
\newcommand{\T}{\mathcal{T}}
\def\one{{\,\hbox{1\kern-.8mm l}}}
\newcommand{\Dslash}{\not{\hbox{\kern-4pt $D$}}}
\newcommand{\pdslash}{\not{\hbox{\kern-2pt $\partial$}}}

\newcommand{\Comment}[1]{{}}

\allowdisplaybreaks

\def\IZ{{\mathbb Z}}

\def\IR{{\mathbb R}}

\newcommand{\bc}{\begin{center}}
\newcommand{\ec}{\end{center}}
\newcommand{\ba}{\begin{array}}
\newcommand{\ea}{\end{array}}
\newcommand{\beq}{\begin{equation}}
\newcommand{\eeq}{\end{equation}}
\newcommand{\bea}{\begin{eqnarray}}
\newcommand{\eea}{\end{eqnarray}}
\newcommand{\bmx}{\begin{pmatrix}}
\newcommand{\emx}{\end{pmatrix}}

\newcommand{\be}{\begin{equation}}
\newcommand{\ee}{\end{equation}}

\def\IB{\relax{\rm I\kern-.18em B}}
\def\IC{{\relax\hbox{\kern.3em{\cmss I}$\kern-.4em{\rm C}$}}}
\def\ID{\relax{\rm I\kern-.18em D}}
\def\IE{\relax{\rm I\kern-.18em E}}
\def\IF{\relax{\rm I\kern-.18em F}}
\def\II{\relax{\rm I\kern-.18em I}}
\def\IZ{\relax{\sf Z\kern-.35em Z}}
\def\Id{\relax{1\kern-.32em 1}}
\def\IG{\relax\hbox{$\inbar\kern-.3em{\rm G}$}}
\def\IR{\relax{\rm I\kern-.18em R}}

\title{Conformal Structure of Massless Scalar Amplitudes Beyond Tree level}

\author{Nabamita Banerjee$^*$, Shamik Banerjee$^\dagger$, Sayali Atul Bhatkar$^*$ and Sachin Jain$^*$}

\affiliation{$^*$Indian Institute of Science Education and Research,\\
Homi Bhabha Rd, Pashan, Pune 411 008, India}
\affiliation{$^\dagger$Institute of Physics,\\  Bhubaneswar, India}

\emailAdd{nabamita@iiserpune.ac.in, banerjeeshamik.phy@gmail.com,
  sayali.bhatkar@students.iiserpune.ac.in, sachin.jain@iiserpune.ac.in }

\abstract{We show that the one-loop on-shell 
four-point scattering amplitude of massless $\phi^4$  scalar field theory in 4D Minkowski
 space time, when Mellin transformed to the Celestial sphere at infinity, transforms covariantly under the 
global conformal group ($SL(2,C)$) on the sphere. The unitarity of the four-point scalar amplitudes
is recast into this Mellin basis. We show that the same 
conformal structure also appears for the two-loop Mellin
amplitude. Finally we comment on some
universal structure for all loop four-point Mellin amplitudes
specific to this theory. }

\preprint{}


\begin{document}

\maketitle

\section{Introduction and Summary}

In a Quantum field theory (QFT), an interesting set of observables are
scattering amplitudes. In flat space they are covariant under the Poincare group. In a four-dimensional QFT on the Minkowski
space, the amplitude are usually constructed out of the asymptotic plane wave solutions to the free wave
equation. The plane waves are eigenstates of the translation
operators. As a result energy-momentum conservation is
manifest in this basis and the amplitudes are translationally
invariant. Whereas Lorentz transformation properties of the 
plane wave states are complicated and hence the whole of SL(2, C) invariance is more subtle. 

Another interesting fact is that the four dimensional Lorentz group $SL(2,\mathbb{C})$  acts
as the group of global conformal transformations on 
the celestial sphere, denoted
by ${\cal{CS}}^2$. This sphere is defined at null infinity where the asymptotic states are specified.
As a result of this, it might be expected that
the scattering amplitude of (massive) massless particles 
should transform in a representation of the global conformal
group, when expressed in terms of right basis states. In particular
for massless particle scattering amplitude, this is  not 
difficult to understand. As we describe in the next
section, in four dimensions, the null momentum of
a massless particle is completely specified by the 
energy and the direction of the three-momentum. Thus modulo a scale, the
null momentum can be completely specified by a
 point on a two dimensional sphere. This two dimensional sphere is a
 space-like cross-section of the light-cone in the 
momentum space. Hence if we think of the light-cone as embedded in
Minkowski space, the two dimensional sphere may be regarded as the
celestial sphere. Thus, when the amplitude is expressed in terms of
the coordinates of this sphere, it is expected to transform covariantly
under the global conformal symmetry. \\

The transformation properties of scattering amplitudes under $SL(2,\mathbb{C})$ was
first considered by Dirac \cite{Dirac:1936fq}.  During last few years, the subject has gotten new
interests, since it has shed some light on the
holographic structure of flat space gravity. We have already said $SL(2,\mathbb{C})$  acts
as the group of global conformal transformations on ${\cal{CS}}^2$. In particular when
gravity is coupled, it has been conjectured that this global
 conformal group gets enhanced to the infinite 
dimensional virasoro algebra. Thus the CFT looks a lot like a
standard 2-D CFT, although the representation of the 
conformal group may be different. Related works on this
topics can be found in \cite{deBoer:2003vf,Banks:2003vp,Barnich:2009se,Barnich:2011ct,Kapec:2014opa,Kapec:2016jld,Cheung:2016iub,He:2014bga,Bianchi:2014gla,Bern:2014oka,Bern:2014vva,Hawking:2016sgy,Cachazo:2014fwa,Strominger:2013lka}. One important ingredient in this study
is the construction of the proper basis states.
Recently in a series of papers \cite{Pasterski:2016qvg,Pasterski:2017kqt,Pasterski:2017ylz,Cardona:2017keg}, Sabrina Pasterski
et.al. have constructed a very interesting integral transform of the
flat space wave functions and
scattering amplitudes of massive and massless particles. In particular
for  massless particle, the transformation takes the form of a Mellin transform
\cite{Pasterski:2017kqt}.  Under this transformation, the momentum-space scattering
amplitude maps to a function, namely the Mellin amplitude, on the Celestial
sphere. The Mellin amplitude transforms like the correlation function of
(quasi -) primary operators of a two dimensional CFT 
defined on the sphere. The putative CFT has operators with all
possible $"$scaling dimensions" of the form $(1+i\lambda)$ 
where $\lambda$ is a real number. The two dimensional spin of an
operator is determined by the helicity of the external particle to which it corresponds to. \\

So far, the conformal structure of flat space tree level massive
scalar amplitudes \cite{Pasterski:2017ylz}  and gluon
amplitudes \cite{Pasterski:2016qvg} have been studied
in the literature. In this paper we use the similar techniques 
to further explore the Mellin transform of the one
and two loop four-point amplitude of a massless scalar 
field theory with $\phi^4$ interaction. Thus, in a sense, this is the first
attempt to understand the conformal structure of flat space scattering
amplitudes beyond tree level. To summarize our main results, we
do see the conformal structure\footnote{Four point function in
a CFT has the form $$\langle \phi(z_1) \phi(z_2) \phi(z_3)
\phi(z_4)\rangle= f (z, \bar z) (\prod_{i<j}^4
|z_{ij}|^{h/3-h_i-h_j}|\bar{z}_{ij}|^{\bar h/3- \bar h_i-\bar
  h_j}),$$ where $f (z, \bar z)$
is some arbitrary function of the cross ratio $(z, \bar z)$ and $z=\frac{z_{12}z_{34}}{z_{13}z_{24}}$.} even at loop level Mellin
amplitudes. In particular, the on-shell one loop four point Mellin amplitude
looks like,
\begin{multline}\label{oneloop}
\tilde{\mathcal{T}}_2=
\frac{i\lambda_R^2}{4}(\frac{2}{\mu})^{-i\Lambda}\bigg[6\pi^3\delta'(\Lambda)
+ \pi^4\delta(\Lambda)   \bigg]{\delta(|z-\bar{z}|)}(\prod_{i<j}^4
|z_{ij}|^{h/3-h_i-h_j}|\bar{z}_{ij}|^{\bar h/3- \bar h_i-\bar
  h_j})[z(z-1)]^{2/3}, \nonumber
\end{multline}
where, $\lambda_R$ is the renormalized coupling constant of the theory
defined at energy scale $\mu$ and $z_i, i=1...,4$ are the position
of each of the four particles on the celestial sphere
${\cal{CS}}^2$. $h_i= \frac{1+ \lambda_i}{2}$ are the conformal dimension of the $i-$th
particle, $\Lambda= \sum \lambda_i$ and $z$ is the conformal cross-ratio function.  
Similar structure holds for two loop amplitude given in equations \eqref{2lpmlin}. The unitarity
property of the QFT amplitude, or equivalently the Optical theorem can also be recasted in terms of the
Mellin amplitude as in \eqref{unitarity}. The results are
subtle and have some universal structure, that leads us to comment on
the form of the amplitude to any arbitrary order in the perturbation
theory for massless $\phi^4$ theory. \\

The paper is organized as follows : In section \ref{sec:sec2}, we
briefly review the conformal basis states for the massive and massless
scalar fields. In the next section \ref{sec:sec3}, we present our main
result, i.e. the one-loop four point Mellin amplitude of the
massless scalar $\phi^4$ theory.  The section is self contained. In section
\ref{sec:sec4}, we explain how the unitarity of the theory can be
implemented on the Mellin space. Finally, in section
\ref{sec:sec5}, we extend the analysis to two-loop amplitude and also
comment on a generic structure for all loop four point amplitude in
this theory. In section \ref{sec:sec6}, we conclude with some
possible future directions to follow.\\

Note: On the day of submission of this draft, another paper \cite{Lam:2017ofc}
appeared on arXiv, that has also addressed the issue of unitarity in
Mellin space.

\section{Conformal Basis for massless scalar fields}
\label{sec:sec2}
In this section we briefly review how four dimensional scattering amplitudes of
a QFT on a flat space can be recasted  with manifest global conformal
symmetry. The construction is given by Sabrina Pasterski
et.al.in \cite{Pasterski:2017kqt} and we refer the readers to their
paper for details. We have
already mentioned that to see the conformal structure of the flat
space amplitudes, we need to first find the right basis for the
asymptotic states that are defined on
the celestial sphere. In \cite{Pasterski:2017kqt}, the authors have defined a new basis
for scalars (massive and massless) and spin one gluons, namely the 
$\textit{conformal primary wave functions}$, that manifests the conformal
structure of their corresponding 4D amplitudes. These conformal 
primaries are characterized by their conformal dimensions and
positions $(w, \bar w)$ on 
a 2-dimensional space, that refer to the 
boundary of the on-shell three diemnsional momentum hyperboloid($H^3$). The
conformal basis for the massless scalars is obtained as a limit of the
corresponding massive one.
More precisely, for a massive particle ($p^0>0$), the 
on-shell momentum hyperboloid looks like
\begin{equation}
(p^0)^2 -(p^1)^2 -(p^2)^2 -(p^3)^2 = m^2,
\end{equation} 
and the corresponding conformal primary wave function defined
on the boundary of this hyperboloid is given as:
\begin{equation}\label{MCP}
\phi_\Delta(x^\mu;w,\bar{w}) = \int_{H^3} \frac{dy}{y^3} dz d\bar{z} \h G_\Delta(y,z,\bar{z};w,\bar{w}) e^{ip_{\mu}(y,z,\bar{z})x^{\mu}}.
\end{equation}
Here, $(y,z,\bar{z})$ are the coordinates on the on-shell momentum hyperboloid
and $(w,\bar{w})$ are the coordinates on the boundary of this
hyperboloid. $G_\Delta$ is the bulk 
to boundary propagator on the hyperboloid and is given as,
\begin{equation}
G_\Delta(y,z,\bar{z};w,\bar{w})= \Big(\frac{y}{y^2+ |z-w|^2}\Big)^{\Delta}.
\end{equation}
This propagator transforms covariantly under the conformal
transformation of the boundary coordinates $w \rightarrow \frac{a
w+b}{c w+d}, \bar w \rightarrow \frac{a \bar w+b}{c \bar w+d} ;
ad-bc=1$. Here, $\Delta$ is a label that defines the conformal
dimension of the propagator. Hence, by construction, the conformal primary
wave function $\phi_\Delta(x^\mu;w,\bar{w})$ has right transformation
properties under conformal transformations of $(w,\bar{w})$.\\
In this paper, we shall be interested in massless scalars. The
corresponding conformal primary can be obtained by appropriately
taking the mass $m \rightarrow 0$ limit of the massive one. For
the massless case, the map is direct as the on-shell momenta are 
null to start with. Hence, they are already at the boundary of the
momentum hyperboloid.
A null momenta can be parametrized as :
\begin{equation}\label{MomRD}
 p = E\left( 1+|z|^2,z+\bar{z},-i(z-\bar{z}),1-|z|^2\right),
 \end{equation}
where, $E$ is an overall scaling of the momenta. 
As shown in \cite{Pasterski:2017kqt}, with the appropriate $m \rightarrow 0$ limit of the bulk to
boundary propagator, the above definition of conformal
primary wave function \eqref{MCP} simply reduces to a 
Mellin transform. Thus for a massless scalar, the conformal primary
takes the form:
\begin{equation}\label{CF1}
\phi(x^\mu;z,\bar{z},\lambda) = \int_0^\infty dE E^{i\lambda} e^{ip(E,z,\bar{z})x}.
\end{equation}
Here, $\lambda$ is a parameter that labels a particular scalar primary
and its implication will be clear in the later section.

\section{4 point one-loop amplitude in massless $\phi^4$ theory}
\label{sec:sec3}

In this paper, we are interested in understanding the conformal
structure of scattering amplitudes at the loop level, for massless
$\phi^4$ theory. To set up our normalizations, the 
theory that we are working with is,
\begin{equation}
S= \int    d^4x [\frac{1}{2}(\partial\phi)^2 - \frac{\hat \lambda}{4!}\phi^4].
\end{equation}

\begin{figure}
\begin{minipage}[t]{.45\linewidth}
\centering
\begin{tikzpicture}[thick,scale=0.8, every node/.style={scale=1.0}]
\begin{feynman}
\vertex(a){\(p_1\)};
\vertex[below = 3.2 cm of a](b){\(p_2\)};
\vertex[above right= 2.2 cm of b](c);
\vertex[above right = 2 cm of c](e){\(p_3\)};
\vertex[below right = 2 cm of c](f){\(p_4\)};

\diagram*{
(a)--[fermion](c),
(b)--[fermion](c),
(e)--[fermion](c),
(f)--[fermion](c)
%
};
\end{feynman}
\end{tikzpicture}
\end{minipage}
\begin{minipage}[t]{.33\linewidth}
\vspace{-2 cm}
+ cross-channel diagrams
\end{minipage}
\caption{Tree level diagram's. Cross-channel corresponds to two other channel of scattering.}
\label{figtree}
\end{figure}
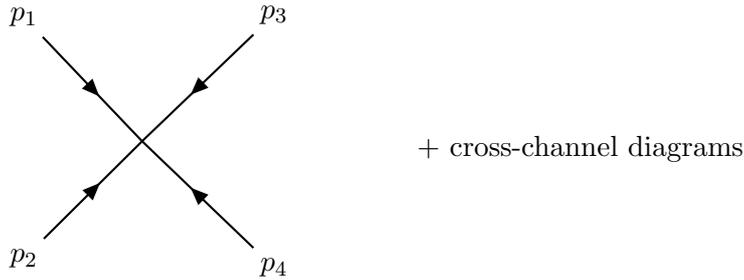

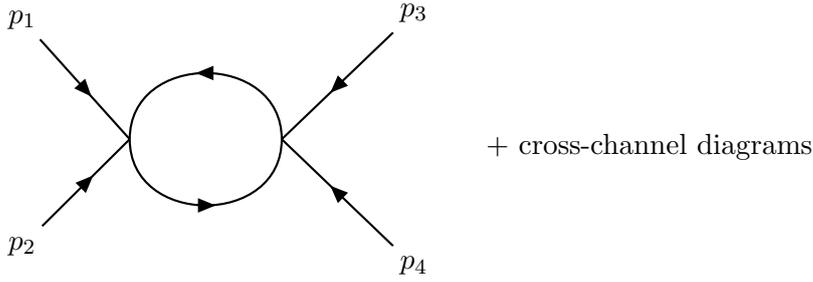
\begin{figure}
\begin{minipage}[t]{.45\linewidth}
\centering
\begin{tikzpicture}[thick,scale=0.8, every node/.style={scale=1.0}]
\begin{feynman}
\vertex(a){\(p_1\)};
\vertex[below = 3 cm of a](b){\(p_2\)};
\vertex[above right= 2 cm of b](c);
\vertex[right = 2 cm of c](d);
\vertex[above right = 2 cm of d](e){\(p_3\)};
\vertex[below right = 2 cm of d](f){\(p_4\)};

\diagram*{
(a)--[fermion](c),
(b)--[fermion](c),
(c)--[fermion, half right](d)--[fermion, half right](c),
(e)--[fermion](d),
(f)--[fermion](d)
%
};
\end{feynman}
\end{tikzpicture}
\end{minipage}
\begin{minipage}[t]{.33\linewidth}
\vspace{-2 cm}
+ cross-channel diagrams
\end{minipage}
\caption{One loop diagram's. Cross-channel corresponds to two other channel of scattering.}
\label{1loop}
\end{figure}
We want to compute the $(2 \rightarrow 2)$ scattering amplitude in this
theory. In this section, we shall only be interested up to one loop
amplitude, although in later sections we shall comment on higher loop
amplitudes. At tree and one loop level, three diagrams (see Fig.\ref{figtree}, Fig.\ref{1loop}) describing three
channels $s, t$ and $u$, contribute to the
$(2 \rightarrow 2)$ scattering amplitude and the non-trivial contribution to the
amplitude is given as,

\begin{equation}\label{AmpM}
A = -i (2\pi)^4 \delta^4 (\Sigma p_i) \bigg[ \lambda_R - \frac{i\pi
  \lambda_R^2}{32\pi^2} + \frac{\lambda{_R}^2}{32\pi^2} 
\big(\ln\frac{s}{\mu^2} + \ln\frac{|t|}{\mu^2} + \ln\frac{|u|}{\mu^2}\big)\bigg].
\end{equation}
Here $\lambda_R$ is the renormalized coupling constant defined at
energy scale $\mu$. $p_1,p_2$ are incoming momenta  and $p_3,p_4$ are  outgoing
momenta. We have defined $(s,t,u)$ as, 
\begin{equation}
s= (p_1 + p_2)^2, \  t=(p_1+p_3)^2, \ u = (p_1+p_4)^2 .
\end{equation}
We want to write the amplitude on the Celestial sphere in an explicit
$s,t$ and $u$ channel symmetric from. For this, we first
rewrite the above amplitude $A(p_i)$ (\ref{AmpM}) as a
function of the on-shell momentum hyperboloid coordinates $(E, z, \bar z)$ 
defined in equation (\ref{MomRD}). Finally, using definition (\ref{CF1}), we Mellin transform this
amplitude to the one defined only on the sphere as,
\begin{equation}
\tilde A(\lambda_i, z_i, \bar z_i)=\prod_{j=1}^4\int_0^{\infty} dE_j
E_j^{i \lambda_i} A(E_i, z_i, \bar z_i).
\end{equation}
 Here $\lambda_i$ are some labels for the Mellin amplitude. The
 inverse transform can be readily obtained as,
\begin{equation}\label{IMellin}
A(E_i, z_i, \bar z_i)=\prod_{j=1}^4\int_{-\infty}^{\infty} \frac{d
  \lambda_j}{2 \pi}
E_i^{-1-i \lambda_j} \tilde A(\lambda_j, z_i, \bar z_i).
\end{equation}
Our construction
closely follows \cite{Pasterski:2017ylz}. The convention is
that all momenta are incoming
with different signs for the energy component. We are now considering
the process where 1 and 2 are incoming and 3 and 4 are outgoing. 
Hence, $$p_1^0=-E_1(1+ |z_1|^2),\ p_2^0=-E_2 (1+ |z_2|^2),\ p_3^0=E_3
(1+ |z_3|^2),\ p_4^0=E_4 (1+ |z_4|^2),$$ and $E_i\ge
0$. Thus, we get,

\begin{equation}
s = 4E_1E_2 |z_{12}|^2, \ t = -4E_1E_3 |z_{13}|^2, \ u = -4 E_1E_4 |z_{14}|^2.
\end{equation}

Using energy momentum conservation relation $p_1 + p_2 +p_3 +p_4 =0$, one
can write three other possible expressions for $s,t$ and $u$ as,

\begin{eqnarray}
s&=& 4E_1E_2 |z_{12}|^2, \  t = -4E_2E_4 |z_{24}|^2, \ u = -4 E_2E_3 |z_{23}|^2 ,\nonumber\\
s&=& 4E_3E_4 |z_{34}|^2, \  t = -4E_1E_3 |z_{13}|^2, \ u = -4
 E_2E_3|z_{23}|^2,\nonumber \\
s&=& 4E_3E_4 |z_{34}|^2, \  t = -4E_2E_4|z_{24}|^2, \ u = -4 E_1E_4 |z_{14}|^2 . \nonumber
\end{eqnarray}
Using the above four expressions for $s,t$ and $u$, we simplify the amplitude in \eqref{AmpM} as, 
\begin{equation}\label{apm1}
A'(E_i, z_i, \bar z_i) = \ln \frac{s|t||u|}{\mu^6}=
\ln\bigg[64 \bigg(\frac{E_1E_2E_3E_4}{\mu^4}\bigg)^{\frac{3}{2}} \prod_{i<j} |z_{ij}|\bigg].
\end{equation}
In writing \eqref{apm1} we have only displayed the momentum dependent
piece of the amplitude and the momentum conserving 
delta function has been omitted. As in
\cite{Pasterski:2017ylz}, it is convenient to change the integration variables 
to an overall frequency
$S\equiv \sum_{i=1}^4 E_i$ and a set of {\it{simplex variables}}
$\sigma_i= S^{-1} E_i \in [0,1]$ that satisfies a constraint
 $\sum_{i=1}^4\sigma_i=1$. Using these new variables we finally get,
\begin{equation}
A'= \ln \bigg[ \frac{S^6}{\mu^6} 64 \bigg(\prod_{i=1}^{4}\sigma_i\bigg)^{\frac{3}{2}} \prod_{i<j} |z_{ij}|\bigg].
\end{equation}
The integral over $E_i$ now gets transformed to integrals over $S$ and
$\sigma_i$, with a proper delta function insertion. The momentum
conserving delta function can be expressed in-terms of these new
simplex variables as, 
\begin{equation}
\delta^4 (\Sigma p_i) = \frac{\delta(|z-\bar{z}|)}{4z_{13}z_{24}\bar{z}_{13}\bar{z}_{24}}\delta(\sigma_1 - \sigma_1^*) \delta(\sigma_2 - \sigma_2^*) \delta(\sigma_3 - \sigma_3^*) \delta(\sigma_4 - \sigma_4^*) ,
\end{equation}
where, we have defined,
\begin{align*}
\sigma_1^* = \frac{z_{24}\bar{z}_{34}}{D z_{12}\bar{z}_{13}}, \sigma_2^* = - \frac{z_{34}\bar{z}_{14}}{D z_{23}\bar{z}_{12}},
\sigma_3^* = - \frac{z_{24}\bar{z}_{14}}{D z_{23}\bar{z}_{13}},\sigma_4^* = \frac{1}{D}, D = 2\Big(\frac{z_{24}\bar{z}_{34}}{ z_{12}\bar{z}_{13}}- \frac{z_{34}\bar{z}_{14}}{ z_{23}\bar{z}_{12}}\Big).
\end{align*}
 After performing the $S$ integral, we get,
\begin{equation}
\int_{0}^{\infty} \frac{dS}{S} S^{i\Lambda} \ln \bigg( \frac{S^6}{\mu^6} 64 \bigg(\prod_{i=1}^{4}\sigma_i\bigg)^{\frac{3}{2}}\left( \prod_{i<j} |z_{ij}|^2\right)^\frac{1}{2} \bigg)
=-6i \delta'(\Lambda) \bigg[2\bigg(\prod_{i=1}^{4}\sigma_i\bigg)^{\frac{1}{4}}\frac{\left( \prod_{i<j} |z_{ij}|^2\right)^\frac{1}{12} }{\mu}\bigg]^{-i\Lambda}.
\end{equation}
Finally, performing the $\sigma$ integrals, we get the one-loop Mellin
amplitude on the sphere as, $\tilde A(\lambda, z, \bar z)=
\tilde{\mathcal{T}_1}+ \tilde{\mathcal{T}_2}$, where, $\tilde{\mathcal{T}_1}$ is the tree level four point
amplitude given as,
\begin{equation}
 \tilde{\mathcal{T}}_1= -\lambda_R(2\pi)^5 \delta(\Sigma_j\lambda_j)\frac{\delta(|z-\bar{z}|)}{4z_{13}z_{24}\bar{z}_{13}\bar{z}_{24}}\Pi_{j=1}^4   (\sigma_*)_j^{i\lambda_j}\nonumber,
\end{equation}
and the one-loop contribution $\tilde{\mathcal{T}_2}$ is given as,
\begin{equation}
\tilde{\mathcal{T}}_2
=(2\pi)^5 \frac{\lambda_R^2}{32\pi^2}\frac{\delta(|z-\bar{z}|)}{4z_{13}z_{24}\bar{z}_{13}\bar{z}_{24}}  (\prod_{j=1}^{4}\sigma_{*j}^{i\lambda_j})\bigg\lbrace 6i\delta'(\Lambda) \bigg[2\bigg(\prod_{i=1}^{4}\sigma_{*i}\bigg)^{\frac{1}{4}}\frac{\left( \prod_{i<j} |z_{ij}|^2\right)^\frac{1}{12} }{\mu}\bigg]^{-i\Lambda} +\delta(\Sigma_j\lambda_j) \pi i\bigg\rbrace.
\end{equation}
The above results can be written in nice conformal covariant form by 
using $ h_i = \bar{h}_i= \frac{1+i\lambda_i}{2}$ and $h= \sum_i h_i$ as:
\begin{equation}\label{tree}
\tilde{\mathcal{T}}_1= -8\pi^5
\lambda_R\delta(\Sigma_j\lambda_j){\delta(|z-\bar{z}|)}(\prod_{i<j}^4
|z_{ij}|^{h/3-h_i-h_j}|\bar{z}_{ij}|^{\bar h/3- \bar h_i-\bar h_j})[z(z-1)]^{2/3} ,
\end{equation}
\begin{multline}\label{oneloop}
\tilde{\mathcal{T}}_2= \frac{i\lambda_R^2}{4}(\frac{2}{\mu})^{-i\Lambda}\bigg[6\pi^3\delta'(\Lambda) + \pi^4\delta(\Lambda)   \bigg]{\delta(|z-\bar{z}|)}(\prod_{i<j}^4   |z_{ij}|^{h/3-h_i-h_j}|\bar{z}_{ij}|^{\bar h/3- \bar h_i-\bar h_j})[z(z-1)]^{2/3}.
\end{multline}
Here we see that the label $\lambda_i$ of the Mellin amplitude is
actually related to the conformal dimensions of each of the four
conformal primaries. The amplitude is channel covariant and transforms
properly as a function of its arguments $(z,
\bar z)$. To be precise, the function takes the form 
in the $s$ channel $12\rightarrow 34$ ($z>1$),
\begin{equation}
f(z,\bar z)=\bigg[z(z-1)\bigg]^{2/3} .
\end{equation}
In the $t$ channel $13\rightarrow 24$ ($0<z<1$)
\begin{equation}
f(z,\bar z)=\bigg[z(1-z)\bigg]^{2/3} .
\end{equation}
In the $u$ channel $14\rightarrow 23$ ($z<0$)
\begin{equation}
f(z,\bar z)=\bigg[z(z-1)\bigg]^{2/3}\footnote{ One can easily
check that the Mellin amplitudes for different channels are related as:
$$\tilde{\mathcal{T}}_{12->34}(z) =
\tilde{\mathcal{T}}_{13->24}(1/z)=\tilde{\mathcal{T}}_{14->23}(1-z),$$ 
with argument$>1$.
The functional form of amplitude is same in all channels only the
ranges of z are different. }.
\end{equation}
Thus, we see that the one-loop amplitude of flat space massless $\phi^4$
theory retains its conformal structure when expressed on the Celestial
sphere. Although this is not a surprise, but, here we also see that
the Function of the cross rations $f(z,\bar z)$ is identical at tree
level and loop-level amplitude. We shall comment more on this
structure in the later section.
\section{Unitarity for Mellin Amplitudes}\label{sec:sec4}

The massless $\phi^4$ theory that we are interested in is an unitary field theory. As we know, in usual
field theories, unitarity of the scattering amplitude implies that the
$S-$matrix satisfies $S S^{\dagger}=1$. Inserting $S= I+ i
\mathcal{T}$, the condition on the non-trivial contribution to the
scattering matrix $\mathcal{T}$ reduces
to, $$-i(\mathcal{T}-\mathcal{T}^\dagger) = 
\mathcal{T} \mathcal{T}^\dagger.$$ This relation plays an extremely
important role in Quantum Field Theory. It computes the imaginary part
in a scattering amplitude and simultaneously predicts that one can
only get the imaginary contribution to scattering when the virtual
particles in a feynman diagram go on-shell. In this section, we would find the
consequence of unitarity on the Mellin amplitudes
$\tilde{\mathcal{T}}$. We shall be presenting the computation for  $s$
channel processes $( 12\rightarrow 34)$. Using the inverse Mellin
transform as defined in \eqref{IMellin}, one can restate the above relation in terms of Mellin
amplitudes. Thus we readily get the following relation,
\begin{align}\label{unitarity}
&\frac{-i}{(2\pi)^4}[\tilde{\T}_{12\rightarrow 34}(\lambda_i, z_i) -
  \tilde{\T}_{34\rightarrow 12}^*(-\lambda_i, z_i)] \nonumber\\
=&\frac{1}{(2\pi)^{8}}\int
   \frac{d^3p}{(2\pi)^{6}}\frac{d^3p'}{4e_p e_{p'}}\int
   d\lambda_pd\lambda_{p'}d\lambda_{P}d\lambda_{P'}
E_p^{-2-i(\lambda_p+\lambda_P)}E_{p'}^{-2-i(\lambda_{p'}+\lambda_{P'})}
 \tilde{\T}_{12\rightarrow p,p'}(\lambda, z)
\tilde{\T}^*_{34\rightarrow P,P'}(\lambda, z)\nonumber\\
=&\frac{4\times 4\pi^2}{(2\pi)^{14}}\int \h dz_p d\bar{z}_p\h dz_{p'} d\bar{z}_{p'}\int d\lambda_pd\lambda_{p'}\tilde{\T}(\lambda_1,\lambda_2,\lambda_p,\lambda_{p'};z_1,z_2,z_p,z_{p'})
\tilde{\T}^*(\lambda_3,\lambda_4,-\lambda_p,-\lambda_{p'};z_3,z_4,z_p,z_{p'}),
\end{align}
Here, ${i=1,2,3,4}$ and we have used following two relations for
simplifications,
\begin{equation}
\begin{split}
d^3p &= 2  E_p e_p\h d E_p\h dz_p d\bar{z}_p,\\
\int E^{-1-i\lambda}dE &= 2\pi\delta(\lambda),~~
e_{p}=E_p \left( 1+|z_p|^2  \right).
\end{split}\end{equation} 

Therefore we see that, for a unitary flat space QFT, the corresponding
Mellin amplitudes has to satisfy \eqref{unitarity}. This relation is
generic and should hold for any unitary QFT. 
We have explicitly checked that the scattering 
amplitude for $ \phi^4$ theory obtained in equations (\ref{tree})
and (\ref{oneloop}) satisfies the above relation. 
For our case the Mellin amplitude\footnote{much like the momentum space amplitude.}
gets its first imaginary contribution at oneloop level and hence the above relation (\ref{unitarity}) gets first
contribution at order $\lambda_R^2$. For the $s$ channel process, using explicit expressions as
given in (\ref{oneloop}), the l.h.s. of (\ref{unitarity}) simplifies to,
\begin{equation}
\frac{-i}{(2\pi)^4}[\tilde{\T}_2(\lambda_i, z_i) -\tilde{\T}_2^*(-\lambda_i, z_i)] 
=\frac{\lambda_R}{8(2\pi)^5}\tilde{\T}_1(\lambda_i, z_i), \quad (i=1,2,3,4),
\end{equation}
where as the rhs of equation (\ref{unitarity}) only picks up the tree level amplitude. Let us rewrite the rhs as,  
\begin{align}
A(z_i)&= 
 \frac{4\times 4\pi^2}{4(2\pi)^{14}}\int \h dz_p d\bar{z}_p\h dz_{p'}
  d\bar{z}_{p'} \nonumber \\
&\int 
d\lambda_pd\lambda_{p'}\tilde{\T}_1(\lambda_1,\lambda_2,\lambda_p,\lambda_{p'};z_1,z_2,z_p,z_{p'})
\tilde{\T}_1(\lambda_3,\lambda_4,-\lambda_p,-\lambda_{p'};z_3,z_4,z_p,z_{p'}).
\end{align}
The last integral is hard to perform in general. Keeping in mind the conformal
structure of the Mellin amplitude, we set the points $z_2 = 1, z_4 =
0, z_1 = \infty$.  This simplifies the computation and we check that :
\begin{equation}
\lim_{z_1\rightarrow\infty}[(\prod_{i<j}^4
|z_{ij}|^{-2h/3+2h_i+2h_j})A(z_1,1,z,0)] =\frac{-1}{2^5} \delta(|z-\bar{z}|)\delta(\Sigma_j\lambda_j)[z(z-1)]^{2/3}.
\end{equation}
The Dirac delta function of $\Lambda$ is easily obtained from two
Dirac delta functions of $\lambda$'s of 
A(by integrating over $\lambda_{p'}$). The other Dirac delta functions
of cross ratios are also handeled in the straight 
forward way. The integral over $\lambda_{p}$ gives : 
\begin{equation}
\int d\lambda_p [\frac{(z \bar{z}_p - \bar{z} z_p)^2 (\bar{z} (-1 + \bar{z}_p) z_p + 
   z (\bar{z} z_p - \bar{z}_p (-1 + \bar{z} + z_p)))^2}{z^2 \bar{z}^2 (\bar{z}_p - 
   z_p)^2 (\bar{z} + z (-1 + \bar{z}_p) - \bar{z}_p + z_p - \bar{z} z_p)^2}]^{i\lambda_p}.
   \end{equation}
It gives a Dirac delta function with three solutions; one of it is
$z=\bar{z}$. The other two solutions are physically non-significant
and hence we discard them. The integrals over remaining variables are
not hard, for example they can be converted to real integrals and be
evaluated using Mathematica. Thus we see that the Mellin amplitudes
nicely satisfies the unitarity constraint of the QFT.

\section{Comments on Higher order Mellin Amplitudes}\label{sec:sec5}
We can extend the above discussion on the conformal structure of the
flat space scattering amplitude to the two and higher loops .  First
we talk about the two loop amplitude. There
are two different types of diagrams that 
contribute at two loop. The first kind is given by diagram in
Fig.\ref{fig2loop}. They lead to momentum dependent log contribution to scattering amplitude given by
\begin{equation}\label{2loop4pt}
T\sim \lambda_R^3\left( \log^2(\frac{s}{\mu})+\log^2(\frac{t}{\mu})+\log^2(\frac{u}{\mu})\right).
\end{equation}
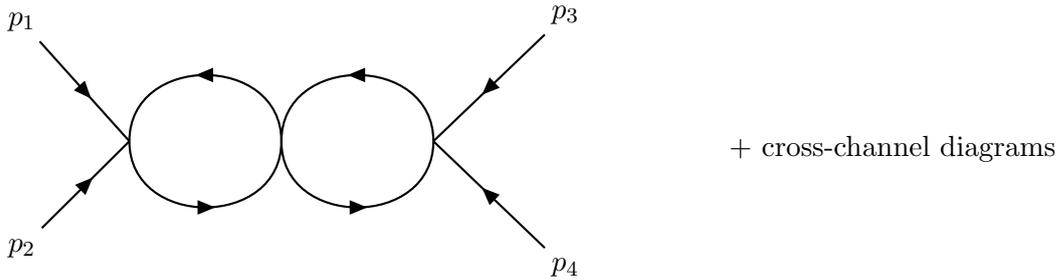
\begin{figure}
\begin{minipage}[t]{.45\linewidth}
\centering
\begin{tikzpicture}[thick,scale=0.8, every node/.style={scale=1.0}]
\begin{feynman}
\vertex(a){\(p_1\)};
\vertex[below = 3 cm of a](b){\(p_2\)};
\vertex[above right= 2 cm of b](c);
\vertex[right = 2 cm of c](d);
\vertex[right = 2 cm of d](g);
\vertex[above right = 2 cm of g](e){\(p_3\)};
\vertex[below right = 2 cm of g](f){\(p_4\)};

\diagram*{
(a)--[fermion](c),
(b)--[fermion](c),
(c)--[fermion, half right](d)--[fermion, half right](c),
(d)--[fermion, half right](g)--[fermion, half right](d),
(e)--[fermion](g),
(f)--[fermion](g)
};
\end{feynman}
\end{tikzpicture}
\end{minipage}
\begin{minipage}[t]{.55\linewidth}
\vspace{-2 cm}
\hspace{2.6 cm}
+ cross-channel diagrams
\end{minipage}
\caption{Leading log two loop diagram}
\label{fig2loop}
\end{figure}

\begin{figure}
\begin{minipage}[t]{.33\linewidth}
\centering
\begin{tikzpicture}[thick,scale=0.8, every node/.style={scale=1.0}]
\begin{feynman}
\vertex(a);
\vertex[left =1 cm of a](b);
\vertex[right=1 cm of a](c);
\vertex[below=1 cm of a](d);
\vertex[above left = 1.5 cm of b](e){\(p_4\)};
\vertex[above right = 1.5 cm of c](f){\(p_3\)};
\vertex[below left = 1.5 cm of d](g){\(p_1\)};
\vertex[below right= 1.5 cm of d](h){\(p_2\)};

\diagram*{
(c)--(b),
(g)--[fermion](d),
(h)--[fermion](d),
(e)--[fermion](b),
(f)--[fermion](c)
};
\end{feynman}
\draw(a)circle(1.22cm);

\end{tikzpicture}
\end{minipage}
\begin{minipage}[t]{.65\linewidth}
\vspace{-2 cm}
\hspace{3 cm}
+ cross-channel diagrams
\end{minipage}
\caption{Two loop contribution.}
\label{fig22loop}
\end{figure}
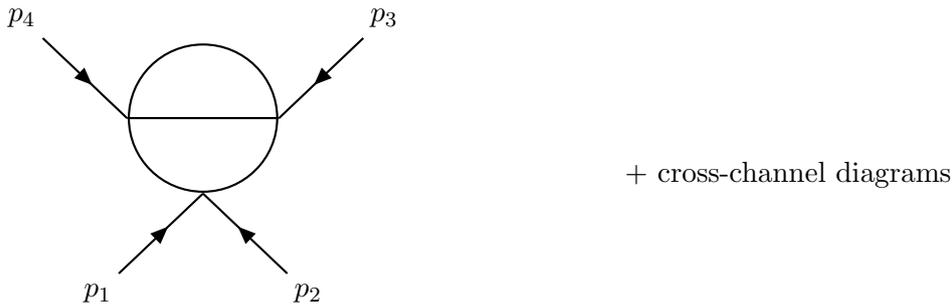

This expression when transformed to Mellin space reduces to :
\begin{align}\label{2lpmlin}
\tilde{\T}_3 \sim
  \lambda_R^3\delta''(\Lambda){\delta(|z-\bar{z}|)}\times(\prod_{i<j}^4
  |z_{ij}|^{h/3-h_i-h_j} |\bar  z_{ij}|^{h/3-h_i-h_j})
\nonumber\\
(\frac{2}{\mu})^{-i\Lambda}[z(z-1)]^{2/3} \lbrace
  z^{-i\frac{\Lambda}{3}}(z-1)^{i\frac{\Lambda}{6}} +
  z^{i\frac{\Lambda}{6}}(z-1)^{i\frac{\Lambda}{6}} 
+ (z-1)^{-i\frac{\Lambda}{3}}z^{i\frac{\Lambda}{6}}\rbrace.
\end{align}
Let us pause here to discuss few universal features of the expression
in \eqref{2lpmlin}. First, we note that, even at the two loop order,
the Mellin amplitude has the proper conformal factor that makes it
transform covariantly under the global conformal
transformation of the boundary sphere. Secondly, the
similar dependence on the form 
of the delta function at tree level, one loop and two loop suggests
that, at $n-$loop level Mellin amplitude, one
should get a delta function dependence of the form 
$\delta^n(\Lambda)$. It is quite interesting that a same cross-ratio
dependent function
$[z(z-1)]^{2/3}$ appeared at tree, one loop and two 
loop level. We expect it to arise at any loop level for leading log term. Also note that
factor in the curly bracket of \eqref{2lpmlin} appears to be universal
at all level. To 
understand why it is universal, let us start with tree level amplitude
$\tilde{\mathcal{T}}_1$ in \eqref{tree}. If we multiply
$\tilde{\mathcal{T}}_1$ in \eqref{tree} by $f(\Lambda)$ defined by :
\begin{equation}\label{sbrt}
f(\Lambda) = \frac{1}{3}\lbrace z^{-i\frac{\Lambda}{3}}(z-1)^{i\frac{\Lambda}{6}} + z^{i\frac{\Lambda}{6}}(z-1)^{i\frac{\Lambda}{6}} + (z-1)^{-i\frac{\Lambda}{3}}z^{i\frac{\Lambda}{6}}\rbrace,
\end{equation}
 we immediately see that the product equates to
 $\tilde{\mathcal{T}}_1$ since $\delta(\Lambda)f(\Lambda) =1$. Thus, we can as well write the tree level Mellin amplitude as $\tilde{\mathcal{T}}_1 f(\Lambda)$. Similarly note
 that the one loop amplitude of \eqref{oneloop} can also be rewritten as
 $\tilde{\mathcal{T}}_2 f(\Lambda) $. The term proportional to
 $\delta(\Lambda)$ in $\tilde{\mathcal{T}}_2$ does not change much like
 the tree level amplitude. The term proportional to the derivative of
 $\delta(\Lambda)$, i.e. 
\begin{equation}
\tilde{\T}_2 \sim \lambda_R^2(\prod_{i<j}^4   |z_{ij}|^{2h/3-2h_i-2h_j})\delta'(\Lambda){\delta(|z-\bar{z}|)}\times
(\frac{2}{\mu})^{-i\Lambda}[z(z-1)]^{2/3} f(\Lambda),
\end{equation}
 can be shown to be identical to the one in \eqref{oneloop}.\footnote{This can be shown by using
$ g(\Lambda)\partial_\Lambda\delta (\Lambda)
 = \partial_\Lambda[\delta (\Lambda)g(0)] - 
\delta(\Lambda)\partial_\Lambda g(\Lambda)$. 
Further when derivative hits $f(\Lambda)$, the terms add up to 0 for $\Lambda = 0$.}
Based on this observation, we can readily extend the result to
arbitrary n-loop order Mellin amplitude. We see that, for any n-loop
order, the momentum space amplitude will have a contribution like :
$\mathcal{M}_{n+1} \sim \lambda_R^{n+1}[log^n(s) + log^n(t) +
log^n(u)] $. The corresponding Mellin amplitude will behave as,
\begin{align}\label{allloop}
\tilde{\T}_{n+1} \sim \lambda_R^{n+1}(\prod_{i<j}^4   |z_{ij}|^{2h/3-2h_i-2h_j})\delta^n(\Lambda){\delta(|z-\bar{z}|)}\times\nonumber\\
2^{-i\Lambda}[z(z-1)]^{2/3} \lbrace
  z^{-i\frac{\Lambda}{3}}(z-1)^{i\frac{\Lambda}{6}} +
  z^{i\frac{\Lambda}{6}}(z-1)^{i\frac{\Lambda}{6}} + 
(z-1)^{-i\frac{\Lambda}{3}}z^{i\frac{\Lambda}{6}}\rbrace.
\end{align}
The numerical coefficient will depend on the exact computation, but
the momentum dependence is fixed as in \eqref{allloop}.\\

The other diagram that contribute at the two loop level is Fig.\ref{fig22loop}.
However, this has no finite momentum dependent contribution to
scattering amplitude (see appendix \ref{appA}). It
only contributes a coupling constant dependent constant
factor. Thus, for the corresponding Mellin amplitude, it contributes similar to \eqref{tree}. The
same feature seems to be true to all loop order Mellin amplitudes, but we do
not have a concrete proof for this yet. \\

\section{Conclusions and Future Directions}\label{sec:sec6}
In this paper, we have studied the conformal structure of the flat
space QFT four-point amplitude of massless $\phi^4$ theory. We have
computed exact results up to two loop order in the perturbation theory
as given in equations \eqref{tree}, \eqref{oneloop} and
\eqref{2lpmlin}. We have also reformulated the role of Unitarity of
QFT for the corresponding Mellin amplitudes.  Equation \eqref{unitarity} is the
constraint that the Mellin amplitudes of any unitary theory has to
satisfy.  In particular, we have shown that the four point Mellin
amplitude of massless  $\phi^4$  theory does satisfy the required
relation. While the conformal structure of the Mellin amplitude
is guaranteed by proper choice of conformal primary wave
functions, the interesting aspect of our results is the universality
of the Mellin amplitude to all these three orders in the perturbation
series. We have seen that the dependence of the Mellin amplitude on the
conformal cross ration factor (the only non-trivial dependence of the
celestial sphere) to
tree, oneloop and two level are identical.  Only difference in them appears as
in the order of derivative of delta function of $\sum\lambda_i$. It is also
certain that similar contribution will be there in all loop Mellin amplitude. 
Based on this observation, we see that all loop answer for $\phi^4$
theory, the leading log part of the Mellin amplitude takes the form :
\begin{align}\label{alllop}
\tilde{\T}_{\rm{all-loop}} \sim (\prod_{i<j}^4
  |z_{ij}|^{2h/3-2h_i-2h_j})\left(\sum_{l=0}^n \lambda_R^{l+1}
 a_l\delta^l(\Lambda)\right){\delta(|z-\bar{z}|)}\times\nonumber\\
(\frac{2}{\mu})^{-i\Lambda}[z(z-1)]^{2/3} \lbrace z^{-i\frac{\Lambda}{3}}(z-1)^{i\frac{\Lambda}{6}} + z^{i\frac{\Lambda}{6}}(z-1)^{i\frac{\Lambda}{6}} + (z-1)^{-i\frac{\Lambda}{3}}z^{i\frac{\Lambda}{6}}\rbrace,
\end{align}
where $\delta^l(\Lambda)$ is $l^{th}$ derivative of the delta function with respect to its argument. For example 
\begin{equation}
\delta^0(\Lambda)=\delta(\Lambda),~~\delta^1(\Lambda)=\delta'(\Lambda),
\end{equation}
and the coefficients $a_l$ are the only undetermined numbers which one
has to compute. It would be remarkable if it turns out that the form in \eqref{alllop} is the entire result for
the four-point  scattering amplitude for $\phi^4$ theory at
all loop.  We have only proved it to be
true up to two loop order. There can be other contributions as well at three and
higher loop order and we do not yet have any concrete comment on
that. 

We end the paper with some comment on possible future
directions.  First of all, it would be nice to find the conformal
covariant amplitudes for QED. For that, one needs to write down the
conformal primaries for asymptotic fermionic states \cite{prog}. This will have
importance on the structure of soft photon theorems of QED. Also, this will provide
us with another example to study the CFT structure of flat space amplitudes.\\  
On a deeper note, in \cite{Agrawal:2016ubh}, the authors outlined how the 4D
scattering amplitudes can be reformulated in the language of a 2D CFT
on the Celestial sphere. To give similar interpretation to our
results, one needs to show how the Mellin amplitudes satisfy all the
properties of a CFT amplitude. In our work we have only shown that
they transform covariantly under the global  conformal
transformation, but a lot is left to do. It would be nice to make this
connection precise. Once the connection is established, the ultimate
goal would be to recast all interesting question of a QFT in their dual 2d CFT
and to compute them directly in the CFT.



\section*{Acknowledgements}
We would like to thank Dileep Jatkar, Rajesh Gopakumar and Shiraz Minwalla for useful
discussions. Shamik B. would like to thank the hospitality at IISER Pune
where the work got started. SJ would like to thank Debjyoti Bardhan
for helping with the graphs. Work of NB is partially supported by a DST/SERB Ramanujan Fellowship.
Finally, we thank the people of India for their generous support for
the basic sciences.

\appendix

\section{Two-loop computation in momentum space}\label{appA}

In this appendix we calculate the contribution of Fig.\ref{fig22loop}
  to the  massless four-point
scattering amplitude at two loop order.
In the following, we will only keep track of terms that are finite and momentum
dependent and not worry about the overall constants. Only the diverging
Gamma functions are explicitly written with other finite ones being omitted. 
Here, $p = p_1+p_2$. The amplitude behave as,
\begin{equation}
\mathcal{M} \sim \int d^dk\int d^dk'
 \frac{1}{k^2(k+p)^2} \frac{1}{k^{'2}(k'+p_3+k)^2}.
\end{equation}
Combining $k'$ dependent factors from the denominator and integrating over them we get, 
\begin{equation}
\mathcal{M} \sim \Gamma(2-d/2)\int_0^1dx\int d^dk \frac{1}{k^2(k+p)^2} \frac{1}{[-x(1-x)(k+p_3)^2]^{2-d/2}}. 
\end{equation}
The x-integral is trivial and momentum independent. 
Combining the three denominators (see for example Kleinert, Chp. 8
\cite{Kleinert:2001ax}) the above expression simplifies to,
\begin{equation}
\mathcal{M}\sim \int_0^1dydz\int d^dk  \frac{z^{1-d/2}}{[(1-y-z)k^2 + y(k+p)^2 + z(k+p_3)^2]^{4-d/2}}. 
\end{equation}
Completing the squares and integrating over k we get,
\begin{equation}
\mathcal{M} \sim \Gamma(4-d)\int_0^1dydz  \frac{z^{1-d/2}}{[yp^2- y^2p^2 - 2yzp p_3]^{4-d}}. 
\end{equation}
 Now, using the relation $p^2 = 2p.p_3$, the integral reduces to,
\begin{equation}
\mathcal{M} \sim \Gamma(\epsilon)\int_0^1dydz \hspace{0.1cm} {z^{-1}}{[y- y^2 - yz ] ^{-\epsilon}}  (p^2)^{-\epsilon}.
\end{equation}
Finally, we have to take $\epsilon\rightarrow 0$ and it gives,
\begin{equation}
\mathcal{M} \sim \int_0^1dz \hspace{0.1cm} {z^{-1}} log(p^2).
\end{equation}
This is the final contribution from  Fig.\ref{fig22loop}. As we see,
the lower limit of the integral is divergent and by usual
renormalization techniques, it cancels with 
diverging pieces of other diagrams. The finite contribution comes from
the upper limit of the Integral and it is simply 0 at the upper
limit. Thus, we conclude that for massless $\phi^4$ theory, the two
loop four point scattering amplitude does not get any momentum
dependence contribution from Fig.\ref{fig22loop}. This feature may not
be true at higher loops.
\bibliographystyle{JHEP}
\bibliography{bms3}

\end{document}